\documentclass[a4paper,12pt]{article}
\usepackage[latin1]{inputenc}
\usepackage{natbib}
\usepackage{amssymb}
\usepackage{amsmath}
\usepackage{amsthm} 
\usepackage{latexsym} 
\usepackage{graphicx}
\usepackage{bm}   
\usepackage{overpic} 
\usepackage[normalem]{ulem}
  
\usepackage{exscale}
\usepackage{amsfonts}
\usepackage[usenames,dvipsnames]{color} 

\def\Om{\Omega}
 \def\ua{\uparrow}
 \def\da{\downarrow}

 \def\wt{\widetilde}

\def\bR{\mathbb{R}}

\def\cC{\mathcal C}

\def\cV{\mathcal V}  

\def\bR{\mathbb R}
\def\cX{\mathcal X}
\def\cF{\mathcal F}
\def\bP{\mathbb P}
\def\bE{\mathbb E}
\def\bN{\mathbb N}

\def\var{\text{var\,}}

\newtheorem{theorem}{Theorem}
\newtheorem{proposition}{Proposition}

\newtheorem{corollary}{Corollary}

\theoremstyle{definition}

\newtheorem{remark}{Remark}

\normalsize
 
\begin{document}
\title{\LARGE\bf  Robust Strategies for Optimal Order Execution \\in the Almgren--Chriss Framework}
\author{\normalsize Alexander
Schied\thanks{Support by Deutsche Forschungsgemeinschaft is gratefully acknowledged.}\\ \normalsize Department of Mathematics\\ \normalsize University of Mannheim\\
\normalsize A5, 6, 68131 Mannheim, Germany\\ 
\normalsize{\tt schied@uni-mannheim.de}
}

\date{\small First version: September 25, 2011\\This version: May 19, 2013}

\maketitle

\begin{abstract}Assuming geometric Brownian motion as unaffected price process $S^0$, \cite{GatheralSchied} derived a strategy for optimal order execution that reacts in a sensible manner on market changes but can still be computed in closed form. Here we will investigate the robustness of this strategy with respect to misspecification of the law of $S^0$. We prove the surprising result that the strategy remains optimal whenever $S^0$ is a square-integrable martingale. We then analyze the optimization criterion of \cite{GatheralSchied} in the case in which $S^0$ is any square-integrable semimartingale and we give a closed-form solution to this problem. As a corollary, we find an explicit solution to the problem of minimizing the expected liquidation costs when the unaffected price process is a square-integrable semimartingale.  The solutions to our problems are found by stochastically solving a finite-fuel control problem without assumptions of Markovianity.   
\end{abstract}
 
\bigskip

\section{Introduction}

This paper can be read from two complementary perspectives. 

From the first perspective, it is a paper on the optimal execution of large orders, which is a problem that was first discussed by \cite{BertsimasLo} and \cite{AlmgrenChriss1,AlmgrenChriss2}. The construction of optimal order execution strategies has considerable practical significance. As always, strategies to be applied in practice should have reasonable quantitative and qualitative properties, should be easy to implement, and, ideally, robust with respect to model misspecification. The first properties are satisfied by the strategy that was derived in \cite{GatheralSchied}. Assuming geometric Brownian motion as unaffected price process $S^0$, this strategy was obtained as the minimizer of a cost functional, which can be regarded as the time-averaged risk of the remaining position. 
In this paper, we investigate how the optimality of this strategy $(x_t^*)$ is affected by changes of the distribution of $S^0$. To this end, we will give a closed-form solution of the strategy that minimizes the cost criterion from \cite{GatheralSchied} if $S^0$ is a general square-integrable semimartingale. Surprisingly, it will turn out that the optimal strategy coincides with $(x_t^*)$ whenever $S^0$ is a square-integrable martingale. In this sense, $(x_t^*)$  is very robust with respect to misspecification of the dynamics of the unaffected price process and, as a consequence, satisfies most requirements one would have on a reasonable order execution strategy.

This brings us to the second perspective, from which this paper can be viewed as a case study in robustness with respect to model uncertainty. The ubiquitous existence of model uncertainty was first emphasized by   \cite{Knight}, but only few systematic approaches to this phenomenon exist to date; we refer to \cite{Cont} and, for an overview over some material, to \cite{FoellmerSchied} and the references therein.  The present paper adds a further particular to the literature on model uncertainty, namely the study of robustness with respect to model misspecification in a stochastic control problem arising in order execution.  This feature of robustness is closely related to the remarkable fact that here it is possible to solve explicitly a stochastic control problem with fuel constraint without assumptions of Markovianity and without using partial differential equations.

As a corollary of our main results, we are able to give an explicit solution to the problem of minimizing the expected liquidation costs in the Almgren--Chriss framework when the unaffected price process is a square-integrable semimartingale. We find that optimal strategies always exist and that the drift enters the corresponding formula in integrated form. One can thus expect that possible misspecifications of the drift may average out.  This relatively stable behavior is in stark contrast to the direct dependence of optimal strategies on the {derivative} of the drift in models with transient price impact as found in \cite{LorenzSchied}. 

In the subsequent Section \ref{Background} we will explain in some detail the background for our study and we will provide a precise formulation of the problem we are looking at. In Section \ref{ResultsSection} we will state our main results. We will start by formulating the results pertaining to martingale dynamics of the unaffected price process. These results are, however, just corollaries of our main result, Theorem \ref{GenSemThm}, which provides the closed-form solution of the optimal strategy  general semimartingale dynamics of $S^0$. As a further corollary, we find an explicit solution to the problem of minimizing the expected liquidation costs when the unaffected price process is a square-integrable semimartingale.  Our formula for the optimal strategy in Theorem \ref{GenSemThm} is obtained by first \emph{guessing} the minimal value  of the optimization problem and then applying a stochastic verification argument to confirm the guess.  In Section \ref{heuristicsSection}, we have included heuristic arguments, based on partial differential equations, which show how this guess can be found. All proofs are provided in Section \ref{ProofsSection}.

\section{Background and problem formulation}\label{Background}

In the continuous-time version of the  market impact model of \cite{AlmgrenChriss1, AlmgrenChriss2} it is assumed that the number of shares in the  portfolio of a trader is described by an absolutely continuous trajectory $t\mapsto x_t$. 
Given this trading trajectory, the price at which transactions occur is
\begin{equation}\label{}
 S^x_t=S^0_t+\eta\dot x_t+{\gamma (x_t-x_0)},
\end{equation}
where $\eta>0$ and $\gamma\ge0$ are constants and $S^0_t$ is the unaffected stock price process. The term $\eta\dot x_t$ corresponds to the \emph{temporary} or \emph{instantaneous impact} of trading $\dot x_t\,dt$ shares at time $t$ and affects only this current order. The term $\gamma (x_t-x_0)$ corresponds to the \emph{permanent price impact} that has been accumulated by all transactions until time $t$. The unaffected price process is usually assumed to be a martingale. There are good reasons, however, why it can make sense to relax the martingale assumption. For instance, there may be other large traders active in the market and their trading activities create a drift on top of random market fluctuations; see e.g., \cite{SchoenebornSchied}.

Let us now consider an \emph{order execution strategy} in which an initial long or short position of $X$ shares is liquidated by time $T$. The asset position of the trader, $(x_t)_{0\le t\le T}$,  thus satisfies the boundary condition $x_0=X$ and $x_T=0$. In such a strategy, $-\dot x_t\,dt$ shares are sold at price $ S^x_t$ at each time $t$. Thus, the \emph{costs} arising from the strategy $(x_t)_{0\le t\le T}$ are
\begin{equation}\label{AC-costsEq}\begin{split}
\cC(x)&:=\int_0^T S^x_t\dot x_t\,dt\\
&=-XS_0-\int_0^Tx_t\,dS^0_t+\eta\int_0^T\dot x_t^2\,dt+\frac\gamma2X^2,
\end{split}\end{equation}
where we have used integration by parts.

The \emph{optimal order execution problem} consists in maximizing a certain objective function, which may involve  revenues and additional risk terms, over a suitable class of admissible trading strategies $(x_t)_{0\le t\le T}$ with side conditions $x_0=X$ and $x_T=0$. The easiest case corresponds to minimizing the expected costs when $S^0$ is a martingale. In this case, the expectation of the stochastic integral $\int_0^Tx_t\,dS^0_t$ vanishes for suitably bounded strategies, and we obtain
\begin{equation}\label{ExpectedExecutionCostsEq}
\bE[\,\cC(x)\,]=-x_0S_0+\frac\gamma2x_0^2+\eta\,\bE\Big[\,\int_0^T\dot x_t^2\,dt\,\Big]
\end{equation}
In this setting, minimization of the expected costs  was first considered in \cite{BertsimasLo} in a discrete-time framework. A simple application of Jensen's inequality shows that the unique  strategy  that minimizes the expected costs \eqref{ExpectedExecutionCostsEq} is characterized by having the  constant trading rate
$$\dot x^{\text{VWAP}}_t=-\frac {x_0}T,
$$
regardless of the particular dynamics  of the martingale $S^0$. When, as is usually assumed in practice, time is parameterized in volume time, such a constant trading rate corresponds to a VWAP strategy, where VWAP stands for volume-weighted average price.

\begin{description}
\item[First problem:] Minimize the expected costs $\bE[\,\cC(x)\,]$ when $S^0$ is not a martingale but a general square-integrable semimartingale. 
\end{description}
Surprisingly, the preceding problem can be solved explicitly in full generality. Our corresponding result, Corollary \ref{expected costs Corollary}, will be derived as a special case of a more general result, Theorem \ref{GenSemThm}. To motivate its statement, we need to look into cost-risk criteria that go beyond the expected costs of an order execution strategy.

 \cite{AlmgrenChriss1,AlmgrenChriss2} were the first to point out that executing orders late in the trading interval $[0,T]$ incurs volatility risk. They therefore suggested to minimize a mean-variance functional  of the form
\begin{equation}\label{meanvariancefunctionaleq}
\bE[\,\cC(x)\,]+\frac\alpha2\var(\cC(x)),
\end{equation}
where $\alpha$ is a risk-aversion parameter. While mean-variance optimization may be appealing to practitioners due to its common use in finance, this approach  has two major disadvantages when applied in order execution. First, 
it is not easy to find mean-variance minimizing strategies unless one restricts strategies to be \emph{deterministic} and assumes that the unaffected price process is a Bachelier model,
\begin{equation}\label{Bachelier}
S^0_t=S_0+\sigma W_t\qquad\text{with $\sigma\neq0$ and $W$ a Brownian motion.}
\end{equation}
In this latter case, calculus of variations easily yields 
\begin{equation}\label{MeanVarOptStraEq}
x^{\text{MV}}_t=X\frac{\sinh\kappa(T-t)}{\sinh\kappa T},\qquad\text{with }\kappa=\sqrt{\frac{\alpha\sigma^2}{2\eta}},
\end{equation}
as the unique deterministic mean-variance optimal strategy. The second disadvantage  stems from the fact that the mean-variance functional \eqref{meanvariancefunctionaleq} is \emph{not time-consistent}, since it involves a squared expectation operator.  As a consequence, an optimal adaptive strategy computed at time $t=0$ loses its optimality at any later time, even if market conditions remain unchanged. Another consequence of time inconsistency is that  techniques from stochastic optimal control cannot be applied directly, which greatly complicates the computation of mean-variance minimizing strategies. We refer to \cite{LorenzAlmgren} and  \cite{Forsyth}.  

The time consistency of the optimization problem can be retained for the maximization of expected utility,
$$\bE[\,U(-\cC(x))\,],
$$
where $U:\bR\to\bR$ is a concave, increasing utility function.  The strategy maximizing the expected utility in the class of all adaptive strategies can be characterized by means of a nonlinear Hamilton--Jacobi--Bellman (HJB) partial differential equation  (PDE) with singular initial condition; see \cite{SchiedSchoeneborn} or \cite{SchoenebornAdaptive}. This equation usually cannot be solved in explicit form unless $U(x)=-e^{-\lambda x}$ and assumption \eqref{Bachelier} holds, in which case we recover \eqref{MeanVarOptStraEq} as optimal strategy; see \cite{SchiedSchoenebornTehranchi}. In all other cases, numerical techniques for solving nonlinear PDEs with singular initial condition will be necessary. Moreover, optimal strategies may have counterintuitive behavior in reaction to certain parameter changes; see \cite[pp. 190-191]{SchiedSchoeneborn}. 

We also refer to \cite{Forsythetal} for another risk criterion that also leads to a singular HJB equation, which is similar to the one found in the maximization of expected utility.

\bigskip

To summarize,  all of the optimization criteria we have discussed so far have at least one of the following four disadvantages:
\begin{itemize}
\item they yield only deterministic strategies that do not react on the movement of asset prices;
\item   they are time-inconsistent;
\item  their computation requires complex numerics for solving  a nonlinear PDE with singular initial condition;
\item or they admit counterintuitive behavior in reaction to certain parameter changes.
\end{itemize}

These properties are all not  desirable from a practical point of view.
\cite{GatheralSchied} therefore proposed another optimization  criterion, which leads to a strategy that is sensitive to changes in the asset price, that can be easily computed in closed form, and whose reaction to parameter changes is completely transparent. 
This optimization criterion  is based on the common practice in risk management to assess the risk of a position of $x>0$ shares as a constant multiple of their current value. Thus, the risk of the asset position $x_t$ at time $t$ is assessed as
$
\wt\lambda x_tS_t^x
$
for some constant $\wt\lambda>0$. This constant $\wt\lambda$ is typically derived from the Value at Risk of a unit asset position under the assumption of  log-normal future returns. As argued in \cite[Remark 2.2]{GatheralSchied}, one could obtain the same formula (but perhaps with a different value of $\wt\lambda$) if Value at Risk is replaced by a coherent risk measure or by any other positively homogeneous risk measure. The optimization criterion proposed in \cite{GatheralSchied} consists in minimizing the following sum of the expected execution costs and the expectation of the time-averaged Value at Risk of the positions $x_t$ held during an order execution strategy $(x_t)$,
$$
\bE\Big[\,\cC(x)+\int_0^T\wt\lambda x_tS^x_t\,dt\,\Big].$$
To simplify notations, we will henceforth consider the minimization of the cost functional
\begin{eqnarray}
\bE\Big[\,\cC(x)+\int_0^T\widetilde\lambda x_t(S^0_t+\gamma x_t)\,dt\,\Big],\label{GSriskEq}
\end{eqnarray}
which can be easily transformed into the minimization of $\bE[\,\cC(x)+\int_0^T\widetilde\lambda x_tS^x_t\,dt\,]$.

In \cite{GatheralSchied}, the minimization of the functional \eqref{GSriskEq} was considered in the case where the unaffected price process $(S^0_t)$ is a risk-neutral geometric Brownian motion, 
\begin{equation}\label{eq:GBM}
S^0_t=S_0e^{\sigma\, W_t-\frac12\,\sigma^2\,t}.
\end{equation}
It was stated in \cite[Theorem 3.2]{GatheralSchied}
 that for $\gamma>0$ the optimal admissible strategy $(x^*_t)$ minimizing the functional \eqref{GSriskEq} within a suitable class of strategies is given by
\begin{equation}\label{GBMoptimalStrategyEq3}
x^*_t={\sinh\big(\nu(T-t)\big)}\bigg[\frac{X}{\sinh\big(\nu T\big)}-\frac{\lambda }{2\nu}\int_0^t\frac{S_s^0}{1+\cosh\big(\nu(T-s)\big)}\,ds\bigg],
\end{equation}
where $\lambda=\wt\lambda/\eta$ and $\nu^2=\wt\lambda\gamma/\eta$.
The optimal strategy in the limiting case $\gamma=\nu=0$ is given by
\begin{equation}\label{GBMoptimalStrategygamma=0Eq2}
x^0_t=\frac{T-t}T\bigg[X-\frac{\lambda T}{4}\int_0^tS_s^0\,ds\bigg],
\end{equation}
see \cite[Theorem 3.1]{GatheralSchied}.

The solutions $(x^*_t)$ and $(x^0_t)$ are clearly adaptive and react  on changes of the asset price. More precisely, they are \emph{aggressive in the money} in the sense that shares are sold faster when stock prices go up. Moreover, in the case $\gamma>0$ we find for the choice $\wt\lambda=\alpha\sigma^2/2\gamma$ that
$$x^*_t=x^{\text{MV}}_t-\frac{\lambda\sinh\big(\nu(T-t)\big)}{2\nu}\int_0^t\frac{S_s^0}{1+\cosh\big(\nu(T-s)\big)}\,ds,
$$
where $x^{\text{MV}}_t$ is as in \eqref{MeanVarOptStraEq}. So $(x_t^*)$ liquidates a given asset position fast than the deterministic mean-variance optimal  strategy $(x^{\text{MV}}_t)$.

\bigskip

A remarkable property of the optimal strategies $(x^*_t)$ and $(x^0_t)$ is that 
they are independent of the volatility $\sigma$ of the unaffected price process $S^0$. This feature already indicates a certain \emph{robustness} of $(x^*_t)$ and $(x^0_t)$ with respect to model uncertainty.  The second goal of this paper is to analyze the robustness of $(x^*_t)$ and $(x^0_t)$ in a systematic manner:

\begin{description}
\item[Second problem:] In setting up our optimization problem \eqref{GSriskEq}, we have assumed that $S^0$ follows a geometric Brownian motion. But suppose that in reality $S^0$ has different dynamics. How will this affect the optimality of our strategies  $(x^*_t)$ and $(x^0_t)$? 
\end{description}
In the next section, we will approach this problem in two steps. In the first step, we will assume that $S^0$ is a  martingale. In this case, the surprising answer to  our question will be that the strategies  $(x^*_t)$ and $(x^0_t)$ remain optimal  whenever $S^0$  is a  rightcontinuous and square-integrable martingale. In this sense, the  optimal strategies  $(x^*_t)$ and $(x^0_t)$ are very robust; their optimality depends only on the martingale property  and not on the particular distribution of  $S^0$.
In the next step, we will also drop the martingale property and assume only that $S^0$ is a square-integrable semimartingale.  By formally taking $\wt\lambda=0$, we will then obtain the solution to our first problem as a special case.

\section{Main results}\label{ResultsSection}

Let us start by formally setting up the optimization problem. All stochastic processes shall be defined on a filtered probability space $(\Om,\cF,(\cF_t),\bP)$ that satisfies the usual conditions and for which $\cF_0$ is $\bP$-trivial, i.e., $\bP[\,A\,]\in\{0,1\}$ for all $A\in\cF_0$. The unaffected price process $S^0$ is assumed to be a  c\`adl\`ag semimartingale that is square-integrable in the sense that
\begin{equation}\label{S0SquareIntEq}
\bE\big[\,\big(\sup_{0\le t\le T}|S^0_t|\big)^2\,\big]<\infty.
\end{equation}

 By $\cX(T,X)$ we denote the class of all admissible strategies for the problem of liquidating $X\ge0$ shares during the time interval $[0,T]$ (analogous  statements hold for the problem of \emph{buying} a position of $X>0$ shares).  This class consists of all  adapted and absolutely continuous strategies that  satisfy the side conditions $x_0=X$ and $x_T=0$ and the integrability condition
\begin{equation}\label{cXConditions}
\bE\Big[\,\int_0^T\dot x_t^2\,dt\,\Big]<\infty.
\end{equation}
  This condition is clearly necessary for  \eqref{GSriskEq}   to make sense and to be finite. In fact, it is  also sufficient:

\bigskip

\begin{proposition}\label{MartCostProp}Under assumption \eqref{S0SquareIntEq}, the cost functional \eqref{GSriskEq}  is well-defined and finite for any $x\in\cX(T,X)$.  If, moreover,  $S^0$ is a martingale, then  the following identity holds:
$$\bE\Big[\,\cC(x)+\int_0^T\widetilde\lambda x_t(S^0_t+\gamma x_t)\,dt\,\Big]=\frac\gamma2X^2-XS_0+\eta\bE\Big[\,\int_0^T\big(\dot x_t^2+\lambda S^0_tx_t+\nu^2x_t^2\big)\,dt\Big],
$$
where again $\lambda=\wt\lambda/\eta$ and $\nu^2=\wt\lambda\gamma/\eta$.
\end{proposition}
\bigskip

We can now state our first result.

\begin{theorem}\label{MartingaleThm}Assume that $S^0$ is {any rightcontinuous and square-integrable martingale} satisfying \eqref{S0SquareIntEq}. For $\nu=\wt\lambda\gamma/\eta>0$,   the unique  strategy minimizing the cost functional
$$\bE\Big[\,\int_0^T\big(\dot x_t^2+\lambda S^0_tx_t+\nu^2x_t^2\big)\,dt\Big]
$$
within $\cX(T,X)$ is given by \eqref{GBMoptimalStrategyEq3}
 and 
 the
value of the minimization problem   is 
\begin{eqnarray}
&&\min_{x\in\cX(T,X)} \bE\Big[\,\int_0^T\big(\dot x_t^2+\lambda S^0_tx_t+\nu^2x_t^2\big)\,dt\Big]=\bE\Big[\,\int_0^T\big((\dot x^*_t)^2+\lambda S^0_tx^*_t+\nu^2(x^*_t)^2\big)\,dt\Big]\nonumber \\
&&=\nu  X^2\coth(\nu  T)+\frac{XS_0}\nu \tanh\Big(\frac{\nu  T}{2}\Big)-\frac1{4{\nu^2} }\bE\Big[\,\int_0^T\Big(S^0_t\tanh\Big(\frac{\nu (T-t)}2\Big)\Big)^2\,dt\,\Big].\label{MartingaleMinimalCostEq}
\end{eqnarray}
For $\nu=0$, the unique optimal strategy is given by
\eqref{GBMoptimalStrategygamma=0Eq2}
and the value of the minimization problem is 
\begin{eqnarray}
\min_{x\in\cX(T,X)} \bE\Big[\,\int_0^T\big(\dot x_t^2+\lambda S^0_tx_t\big)\,dt\Big]&=&\bE\Big[\,\int_0^T\big((\dot x^0_t)^2+\lambda S^0_tx^0_t\big)\,dt\Big]\label{MartingaleMinimalCostgamma=0Eq}\\
&=&\frac  {X^2}T+\frac12{XS_0}T-\frac1{16 }\bE\Big[\,\int_0^T\big(S^0_t (T-t)\big)^2\,dt\,\Big].\nonumber
\end{eqnarray}
\end{theorem}

\bigskip

The strategies \eqref{GBMoptimalStrategyEq3} and \eqref{GBMoptimalStrategygamma=0Eq2} are independent of the particular law of $S^0$ whenever $S^0$ is a martingale. This has the immediate consequence that these strategies also minimize the following \emph{robust} cost functionals.

\bigskip

\begin{corollary}Let $\mathcal Q$ be any set of equivalent probability measures $Q$ on $(\Om,\cF)$ under which the stochastic process $S^0$ is a square-integrable martingale satisfying \eqref{S0SquareIntEq} and for which $(\Om,\cF,(\cF_t),Q)$ satisfies the usual conditions.
When $\nu=\wt\lambda\gamma/\eta>0$, the  strategy \eqref{GBMoptimalStrategyEq3} minimizes the   cost functional
\begin{equation}\label{risk measure cost fctl}
\sup_{Q\in\mathcal Q}\bE_Q\Big[\,\int_0^T\big(\dot x_t^2+\lambda S^0_tx_t+\nu^2x_t^2\big)\,dt\Big],
\end{equation}
and the minimal cost is given by 
\begin{eqnarray*}\lefteqn{\min_{x\in\cX(T,X)}\sup_{Q\in\mathcal Q}\bE_Q\Big[\,\int_0^T\big(\dot x_t^2+\lambda S^0_tx_t+\nu^2x_t^2\big)\,dt\Big]=\sup_{Q\in\mathcal Q}\bE_Q\Big[\,\int_0^T\big(\dot (x^*_t)^2+\lambda S^0_tx^*_t+\nu^2(x^*_t)^2\big)\,dt\Big]}\\
&=&\nu  X^2\coth(\nu  T)+\frac{XS_0}\nu \tanh\Big(\frac{\nu  T}{2}\Big)-\frac1{4{\nu^2} }\inf_{Q\in\mathcal Q} \bE_Q\Big[\,\int_0^T\Big(S^0_t\tanh\Big(\frac{\nu (T-t)}2\Big)\Big)^2\,dt\,\Big].
\end{eqnarray*}
An analogous statement holds in case $\nu=0$.

\end{corollary}

\bigskip

\begin{remark}Let us summarize the positive and negative properties of the  optimal strategies \eqref{GBMoptimalStrategyEq3} and \eqref{GBMoptimalStrategygamma=0Eq2}.
\begin{itemize}
\item[+] As shown by Theorem \ref{MartingaleThm}, the optimal strategies have a remarkable robustness property: they minimize the cost functional \eqref{GSriskEq} (or its robust version \eqref{risk measure cost fctl}) for every martingale $S^0$, regardless of the specific law of that process. If one accepts the cost criterion \eqref{GSriskEq}  and the assumption that the unaffected price process is well-described by martingale dynamics, then  the strategies \eqref{GBMoptimalStrategyEq3} and \eqref{GBMoptimalStrategygamma=0Eq2} will be optimal even in a situation of model uncertainty. These strategies are therefore robust with respect to model risk.
 \item[+]  The strategies \eqref{GBMoptimalStrategyEq3} and \eqref{GBMoptimalStrategygamma=0Eq2} are given in explicit form and can be very easily implemented in practice, without the need for complex numerical methods. Their dependence on asset prices and model parameters is completely transparent. Also in this sense these strategies are very robust. 
 \item[-] As a disadvantage, it should be noted that the strategies \eqref{GBMoptimalStrategyEq3} and \eqref{GBMoptimalStrategygamma=0Eq2} can become negative. In practice, this will just lead to the early termination of the strategy and to the early liquidation of the asset position. But, as a consequence, the strategy will lose its optimality property in such a scenario. As discussed in \cite[Section 4]{GatheralSchied}, the probability that strategies  become negative will be very small with reasonable parameter choices. The possible negativity of strategies can thus be seen as an effect that may be as negligible as the possible negativity of the unaffected price process in the Bachelier model, which is frequently employed in order execution. \hfill$\diamondsuit$
\end{itemize}
\end{remark}

\bigskip

Theorem \ref{MartingaleThm} is in fact a corollary of our next result, which applies to the situation of a general semimartingale $S^0$. In this case, the optimal strategy can be conveniently written in terms of stochastic integrals with respect to the  semimartingale
\begin{equation}\label{YEq}
Y_t:=-\frac1\eta (S^0_t-S_0)+\lambda\int_0^tS^0_s\,ds.
\end{equation}
To be precise, stochastic integrals starting in $t>0$ will be defined as
$$\int_t^T\xi_u\,dY_u:=\int_{t+}^T\xi_u\,dY_u=\int_0^T\xi_u\,dY_u-\int_0^t\xi_u\,dY_u,
$$
and stochastic processes of conditional expectations such as
$$\bE\Big[\,\int_{t}^{T}\sinh(\nu (T-u) )\,dY_u\,\Big|\,\cF_t\,\Big],\qquad 0\le t\le T,
$$
will be understood as the right-continuous version of this process, which exists since our underlying probability space satisfies the usual conditions.

\bigskip

\begin{theorem}\label{GenSemThm}For a general  semimartingale $S^0$ satisfying \eqref{S0SquareIntEq}, there exists a unique strategy in $\cX(T,X)$ that minimizes the functional $\bE[\,\cC(x)+\int_0^T\widetilde\lambda x_t(S^0_t+\gamma x_t)\,dt\,]$ within the class $\cX(T,X)$. For $\nu=\wt\lambda\gamma/\eta>0$, this strategy is given by
\begin{eqnarray}\label{Thm2OptStrategyEq}
x^*_t
={\sinh\big(\nu (T-t)\big)}\bigg[\frac{X}{\sinh\big(\nu  T\big)}-\frac12\int_0^t\frac{\bE\big[\,\int_{s}^{T}\sinh(\nu (T-u) )\,dY_u\,\big|\,\cF_s\,\big]}{\big(\sinh\big(\nu (T-s)\big)\big)^2}\,ds\bigg],
\end{eqnarray}
and the value of the minimization problem is
\begin{eqnarray}\lefteqn{\min_{x\in\cX(T,X)}\bE\Big[\,\cC(x)+\int_0^T\widetilde\lambda x_t(S^0_t+\gamma x_t)\,dt\,\Big]=\bE\Big[\,\cC(x^*)+\int_0^T\widetilde\lambda x^*_t(S^0_t+\gamma x^*_t)\,dt\,\Big]}\nonumber
\\&&=\frac\gamma2X^2-XS_0+\eta\bigg(\nu  X^2\coth(\nu  T)+\frac X{\sinh(\nu  T)}\bE\Big[\,\int_0^{T} \sinh(\nu (T-u) )\,dY_u\,\Big]\label{CostFunctional}\\
&& \qquad-\frac14\bE\bigg[\,\int_0^T\Big(\frac{\bE\big[\,\int_{t}^{T}\sinh(\nu (T-u) )\,dY_u\,\big|\,\cF_t\,\big]}{\sinh(\nu  (T-t))}\Big)^2\,dt\bigg]\bigg).\nonumber
\end{eqnarray}
For $\nu=0$, the optimal strategy is 
\begin{equation}\label{Thm2OptStrategynu=0Eq}
x^0_t=\frac{T-t}{T}\bigg(X-\frac12\int_0^t\frac T{(T-s)^2}\bE\Big[\,\int_s^T(T-u)\,dY_u\,\Big|\,\cF_s\,\Big]\,ds\bigg),
\end{equation}
and the value of the optimization problem is given by
\begin{eqnarray}\lefteqn{\min_{x\in\cX(T,X)}\bE\Big[\,\cC(x)+\int_0^T\widetilde\lambda x_t(S^0_t+\gamma x_t)\,dt\,\Big]=\bE\Big[\,\cC(x^0)+\int_0^T\widetilde\lambda x^0_t(S^0_t+\gamma x^0_t)\,dt\,\Big]}\nonumber
\\&&=-XS_0+\eta\bigg(\frac{ X^2}T+\frac X{T}\bE\Big[\,\int_0^{T} (T-u) \,dY_u\,\Big]-\frac14\bE\bigg[\,\int_0^T\Big(\frac{\bE\big[\,\int_{t}^{T}(T-u) \,dY_u\,\big|\,\cF_t\,\big]}{T-t}\Big)^2\,dt\bigg]\bigg).\nonumber
\end{eqnarray}
\end{theorem}

\bigskip

In the proof of Theorem \ref{GenSemThm} it is actually shown that the respective strategies \eqref{Thm2OptStrategyEq}  or \eqref{Thm2OptStrategynu=0Eq} minimize any cost functional of the form
$$\bE\Big[\,\int_0^Tx_t\,dY_t+\int_0^T(\dot x_t^2+\nu^2x_t^2)\,dt \,\Big],
$$
where $Y$ is an arbitrary semimartingale with
$$\bE\big[\,\big(\sup_{0\le t\le T}|Y_t|\big)^2\,\big]<\infty.
$$
The specific form \eqref{YEq} of $Y$ is not needed in the proof. Taking $Y:=\frac1\eta S^0$ and $\nu=0$ thus yields the following corollary.

\bigskip

\begin{corollary}\label{expected costs Corollary}For a general  semimartingale $S^0$ satisfying \eqref{S0SquareIntEq}, there exists a unique strategy in $\cX(T,X)$ that minimizes the expected costs $\bE[\,\cC(x)\,]$  within the class $\cX(T,X)$. This strategy is given by
\begin{equation}\label{expected costs optimal strategy}
x^0_t=\frac{T-t}{T}\bigg(X-\frac1{2\eta}\int_0^t\frac T{(T-s)^2}\bE\Big[\,\int_s^T(T-u)\,dS^0_u\,\Big|\,\cF_s\,\Big]\,ds\bigg),
\end{equation}
and the value of the optimization problem is given by
\begin{eqnarray}\lefteqn{\min_{x\in\cX(T,X)}\bE\Big[\,\cC(x)\,\Big]=\bE\Big[\,\cC(x^0)\,\Big]=\frac\gamma2X^2-XS_0}\nonumber
\\&&+\eta\frac{ X^2}T+\frac X{T}\bE\Big[\,\int_0^{T} (T-u) \,dS^0_u\,\Big]-\frac1{4\eta}\bE\bigg[\,\int_0^T\Big(\frac{\bE\big[\,\int_{t}^{T}(T-u) \,dS^0_u\,\big|\,\cF_t\,\big]}{T-t}\Big)^2\,dt\bigg].\nonumber
\end{eqnarray}
\end{corollary}

\bigskip

Let $S^0=S_0+M+A$ be the decomposition of the semimartingale $S^0$ into a local martingale $M$ and an adapted process $A$ of locally finite variation. Let us assume for simplicity that $A$ has integrable total variation over $[0,T]$. Then \eqref{S0SquareIntEq} implies that $M$ is uniformly integrable over $[0,T]$ and thus a true martingale.  Hence, the optimal strategy \eqref{expected costs optimal strategy} can be written as
$$x^0_t=\frac{T-t}{T}\bigg(X-\frac1{2\eta}\int_0^t\frac T{(T-s)^2}\bE\Big[\,\int_s^T(T-u)\,dA_u\,\Big|\,\cF_s\,\Big]\,ds\bigg).
$$
Here the drift $A$ enters the optimal strategy $x^0$ basically in integrated form, and so one can expect that possible misspecifications of the drift may average out.  This relatively stable behavior is in stark contrast to the direct dependence of optimal strategies on the \emph{derivative} of the drift in models with transient price impact as found in \cite{LorenzSchied}.

We conclude this section with the following  outlook on possible generalizations of our problem.

\begin{remark}[Ambiguity with respect to market impact parameters]Our approach yields robustness with respect to the law of $S^0$ but it requires that the exact values of the market impact parameters, $\gamma$ and $\eta$, are known.  In reality, these  parameters will be uncertain to some degree as well, but our results say nothing about the robustness with respect to these parameters. Also, our results rely in an essential way on the assumption that temporary impact is linear. It is not known to the author how the optimization problem from \cite{GatheralSchied} can be solved for nonlinear temporary impact.  Nonlinear price impact is often observed in data of financial transactions; see, e.g., \cite{AlmgrenHauptmanLi}. It also arises in a natural way when one attempts to solve our minimization problem under the additional \lq no-buy\rq\ constraint $\dot x_t\le0$ for $t\in[0,T]$. \hfill$\diamondsuit$ \end{remark}

\begin{remark}[Multiplicative market impact model]One of the shortcomings of the Almgren--Chriss model is that asset prices can become negative when price impact gets too large. To avoid negative prices,
\cite{BertsimasLo}  proposed the following multiplicative price impact model,
$$\wt S^x_t:=S^0_t\exp\big(\gamma x_t+\eta\dot x_t\big),
$$
where $\gamma,\eta\ge0$. Naturally, this model goes well along with geometric Brownian motion as unaffected price process $S^0$; see also \cite{Forsyth} and \cite{Forsythetal}. When defining $\wt \cC(x)=\int_0^T\wt S^x_t\dot x_t\,dt$, one is led to the minimization of the functional
\begin{equation}\label{mult functional}
\bE\Big[\,\wt\cC(x)+\wt\lambda\int_0^T\wt S^x_tx_t\,dt\,\Big].
\end{equation}
 When $S^0$ is a martingale, this cost functional can be expressed as
$$\bE\Big[\,S^0_T\int_0^T\exp\big(\gamma x_t+\eta\dot x_t\big)(\dot x_t+\wt\lambda x_t)\,dt\,\Big].
$$
Its minimization therefore boils down to the minimization of the classical \lq action functional\rq\  $\int_0^T\exp\big(\gamma x_t+\eta\dot x_t\big)(\dot x_t+\wt\lambda x_t)\,dt$ over \emph{deterministic} strategies  $x\in\cX(T,X)$; see also Remark 4.1 in \cite{GatheralSchied}. When $S^0$ is not a martingale, however, the solution to the problem of minimizing \eqref{mult functional} is not known to the author.
\hfill$\diamondsuit$ \end{remark}

\section{Heuristic derivation of the solution}\label{heuristicsSection}

The solution to our optimization problems is obtained by \emph{guessing} the formulas \eqref{Thm2OptStrategyEq} and \eqref{CostFunctional} for the optimal strategy and value 
of the minimization problem and by applying  stochastic verification arguments to show that these guesses are correct. In this section, we explain how the formulas \eqref{Thm2OptStrategyEq} and \eqref{CostFunctional} can be guessed heuristically. 
To this end, we assume that $S:=S^0$ is a diffusion process with dynamics
$$dS_t=\sigma(S_t)\,dW_t+b(S_t)\,dt,
$$
with sufficiently bounded and regular coefficients $\sigma(\cdot)$ and $b(\cdot)$. Then, 
\eqref{AC-costsEq} implies that
\begin{eqnarray*}\cC(x)&=&\frac\gamma2X^2-XS_0-\int_0^Tx_t\,dS_t+\eta\int_0^T\dot x_t^2\,dt\\
&=&\frac\gamma2X^2-XS_0-\int_0^Tx_t\sigma(S_t)\,dW_t-\int_0^Tx_tb(S_t)\,dt+\eta\int_0^T\dot x_t^2\,dt.
\end{eqnarray*}
Hence, if $x\in\cX(T,X)$ is sufficiently bounded,
\begin{eqnarray*}\bE\Big[\,\cC(x)+\int_0^T\widetilde\lambda x_t(S^0_t+\gamma x_t)\,dt\,\Big]=\frac\gamma2X^2-XS_0+\eta \bE\Big[\,\int_0^T\Big(\dot x_t^2+ x_t\big(\lambda S_t-\wt b(S_t)\big)+\nu^2x_t^2\Big)\,dt\,\Big],
\end{eqnarray*}
where $\wt b(x)=\eta^{-1}b(x)$. We thus define the value function 
$$C(T,X,S):=\inf_x \bE\Big[\,\int_0^T\Big(\dot x_t^2+ x_t\big(\lambda S_t-\wt b(S_t)\big)+\nu^2x_t^2\Big)\,dt\,\Big],$$
where the infimum is taken over all strategies $x\in\cX(T,X)$ for which $\int_0^Tx_t\sigma(S_t)\,dW_t$ is a martingale (by \eqref{xt2EstimateEq} below this is actually the case for \emph{all} $x\in\cX(X,T)$ as soon as $\sigma(\cdot)$ is bounded). 
Now we parameterize these strategies $x\in\cX(T,X)$ by their derivative, $v_t=\dot x_t$, and let $\cV(T,X)$ denote the corresponding set of controls. For $v\in\cV(T,X)$ we define $x^v_t:=X+\int_0^t v_s\,ds$. Then $C(T,X,S)$ can be written as
$$C(T,X,S)=\inf_{v\in\cV(T,X)} \bE\Big[\,\int_0^T\Big(v_t^2+ x^v_t\big(\lambda S_t-\wt b(S_t)\big)+\nu^2(x^v_t)^2\Big)\,dt\,\Big].
$$
Dynamic programming suggests that 
$$C(T-t,x^v_t,S_t)+\int_0^t\Big(v_s^2\,ds+x^v_s\big(\lambda S_s-\wt b(S_s)\big)+\nu^2(x^v_s)^2\Big)\,ds
$$
should be a submartingale for every $v\in\cV(T,X)$ and a martingale as soon as $v$ is optimal. Let us assume that $C$ is  smooth. Then an application of It\^o's formula suggests that $C$ should solve the following degenerate quasilinear PDE:
\begin{equation}\label{HJBeq2}
C_T=\frac12\sigma^2(S)C_{SS}+ b(S)C_S+ X(\lambda S-\wt b(S))+\nu^2 X^2+\inf_{v\in\bR}(v^2+vC_X).
\end{equation}
In addition, the {fuel constraint} $\int_0^Tv_t\,dt=-X$ required from strategies in $\cV(T,X)$ suggests that the value function $C$ should satisfy a singular initial condition of the form
\begin{equation}\label{InitialCond}
\lim_{T\da0}C(T,X,S)=\begin{cases}0&\text{if $X=0$,}\\
+\infty&\text{if $X\neq0$.}
\end{cases}
\end{equation}
The intuitive explanation for this initial condition is that a nonzero asset position with no time left for its liquidation means that the liquidation constraint has been violated. Excluding this violation requires an infinite penalty; see also \cite{SchiedSchoenebornTehranchi} for similar  effects in utility maximization for order execution.

To solve \eqref{HJBeq2}, \eqref{InitialCond}, we make the ansatz
$$C(T,X,S)=\nu  X^2\coth(\nu  T)+Xh(T,S)+g(T,S).
$$
Then
\begin{eqnarray*}
C_T&=&-\frac{\nu ^2X^2}{(\sinh(\nu  T))^2}+Xh_T(T,S)+g_T(T,S)\\
C_X&=&2\nu  X\coth(\nu  T)+h(T,S)\\
C_S&=&Xh_S(T,S)+g_S(T,S)\\
C_{SS}&=&Xh_{SS}(T,S)+g_{SS}(T,S).
\end{eqnarray*}
Plugging this into \eqref{HJBeq2} yields
\begin{eqnarray*}
\lefteqn{-\frac{\nu ^2X^2}{(\sinh(\nu  T))^2}+Xh_T(T,S)+g_T(T,S)}\\
&&=\frac12\sigma^2\Big(Xh_{SS}(T,S)+g_{SS}(T,S)\Big)+bXh_S(T,S)+bg_S(T,S)\\
&&\qquad +X(\lambda S-\wt b(S))+\nu ^2X^2-\frac14\Big(2\nu  X\coth(\nu  T)+h(T,S)\Big)^2.
\end{eqnarray*}
Since 
\begin{eqnarray*}
\lefteqn{\nu ^2X^2-\frac14\Big(2\nu  X\coth(\nu  T)+h(T,S)\Big)^2}\\
&&=-\frac{\nu ^2X^2}{(\sinh(\nu  T))^2}-\nu  X\coth(\nu  T)h(T,S)-\frac14 h(T,S)^2,
\end{eqnarray*}
we obtain
\begin{eqnarray}\lefteqn{X h_T(T,S)+g_T(T,S)}\nonumber\\
&&=\frac12\sigma^2\Big(X h_{SS}(T,S)+g_{SS}(T,S)\Big)+bX h_S(T,S)+bg_S(T,S)\label{HeuristicPDEs}\\
&&\qquad +X(\lambda S-\wt b(S))-\nu  X\coth(\nu  T)h(T,S)-\frac14h(T,S)^2.\nonumber
\end{eqnarray}
Equating all terms containing $X$ yields
\begin{eqnarray*}h_T=\frac12\sigma^2h_{SS}+bh_S+\lambda S-\wt b(S)-\nu  \coth(\nu  T)h.
\end{eqnarray*}
As initial condition we take $h(0,S)=0$. This initial-value problem can be solved by means of the Feynman-Kac formula \cite[Theorem 5.7.6]{KaratzasShreve}. To this end, we define $\hat h(t,x):=h(T-t,x)$ for some fixed $T>0$. Then $\hat h$ satisfies the terminal condition $\hat h(T,x)=0$ and the PDE
$$-\hat h_t+\nu  \coth(\nu  (T-t))\hat h=\frac12\sigma^2\hat h_{SS}+b\hat h_S+\lambda S-\wt b(S).
$$
It follows from the quoted Feynman-Kac formula that, for $t<T$ and $Y$ is as in \eqref{YEq},
\begin{eqnarray*}
\hat h(t,x)&=&\bE\Big[\,\int_t^{T}\big(\lambda S_{u}-\wt b(S_u)\big)\exp\Big(-\int_t^u\nu  \coth(\nu  (T-r))\,dr\Big)\,du\,\Big|\,S_t=x\,\Big]\\
&=&\bE\Big[\,\int_t^{T}\exp\Big(-\int_t^u\nu  \coth(\nu  (T-r))\,dr\Big)\,dY_u\,\Big|\,S_t=x\,\Big]\\
&=&\frac1{\sinh(\nu  (T-t))}\bE\Big[\,\int_t^{T}\sinh(\nu (T-u) )\,dY_u\,\Big|\,S_t=x\,\Big].
\end{eqnarray*}
Hence,
$$h(T-t,S_t)=\hat h(t,S_t)=\frac1{\sinh(\nu  (T-t))}\bE\Big[\,\int_t^{T}\sinh(\nu (T-u) )\,dY_u\,\big|\,\cF_t\,\Big],
$$
and in particular
$$h(T,S_0)=\hat h(0,S_0)=\frac1{\sinh(\nu  T)}\bE\Big[\,\int_0^{T} \sinh(\nu (T-u) )\,dY_u\,\Big].
$$

The PDE for the function $g$ is obtained by taking $X=0$ in \eqref{HeuristicPDEs}:
$$g_T=\frac12\sigma^2g_{SS}+bg_S-\frac14h(T,S)^2.
$$
The initial condition must be $g(0,S)=0$. It follows that
\begin{eqnarray*}
g(T,S_0)&=&-\bE\bigg[\,\int_0^T\frac14\big(h(T-t,S_t)\big)^2\,dt\,\bigg]\\
&=&-\frac14\bE\bigg[\,\int_0^T\bigg(\frac{\bE\big[\,\int_t^{T}\sinh(\nu (T-u) )\,dY_u\,\big|\,\cF_t\,\big]}{\sinh(\nu  (T-t))}\bigg)^2\,dt\bigg].
\end{eqnarray*}
Putting everything together, we obtain
\begin{eqnarray}C(T,X,S_0)&=&\nu  X^2\coth(\nu  T)+\frac X{\sinh(\nu  T)}\bE\Big[\,\int_0^{T} \sinh(\nu (T-u) )\,dY_u\,\Big]\nonumber\\
&& -\frac14\bE\bigg[\,\int_0^T\bigg(\frac{\bE\big[\,\int_t^{T}\sinh(\nu (T-u) )\,dY_u\,\big|\,\cF_t\,\big]}{\sinh(\nu  (T-t))}\bigg)^2\,dt\bigg]\label{PermImpCostEq}.
\end{eqnarray}
Note that this formula is independent of the particular characteristics $\sigma(\cdot)$ and $b(\cdot)$ of the dynamics of $S$. We therefore may guess that the formula remains true for even more general semimartingales, which lead us to asserting \eqref{CostFunctional}. 

Standard arguments in control suggest that the optimal strategy $x^*$ is defined through that $v^*$ that attains the infimum in \eqref{HJBeq2}. More precisely, 
$x^*$ should be the solution of the ODE
\begin{eqnarray*}\dot x^*_t&=&-\frac12C_X(T-t,x^*_t,S_t)=-\nu  x^*_t\coth(\nu (T-t))-\frac12h(T-t,S_t) .
\end{eqnarray*}
This ODE is solved by
\begin{eqnarray}x^*_t&=&{\sinh\big((T-t)\nu \big)}\bigg[\frac{X}{\sinh\big(\nu  T\big)}-\frac12\int_0^t\frac{h(T-s,S_s)}{\sinh\big(\nu (T-s)\big)}\,ds\bigg]\nonumber\\
&=&{\sinh\big((T-t)\nu \big)}\bigg[\frac{X}{\sinh\big(\nu  T\big)}-\frac12\int_0^t\frac{\bE\big[\,\int_s^{T}\sinh(\nu (T-u) )\,dY_u\,\big|\,\cF_s\,\big]}{\big(\sinh\big(\nu (T-s)\big)\big)^2}\,ds\bigg],\nonumber
\end{eqnarray}
and this formula suggested the assertion  \eqref{Thm2OptStrategyEq}.

\section{Proofs}\label{ProofsSection}

\bigskip\noindent{\bf Proof of Proposition \ref{MartCostProp}:}  Let $x\in\cX(T,X)$ be given. We have 
$$\cC(x)=\int_0^T S^x_t\dot x_t\,dt=\int_0^TS^0_t\dot x_t\,dt+\eta\int_0^T\dot x_t^2\,dt+\gamma\int_0^T (x_t-x_0)\dot x_t\,dt.
$$
The rightmost integral is equal to 
$$\gamma x_0^2+\gamma\int_0^Tx_t\dot x_t\,dt=\frac\gamma2x_0^2=\frac\gamma2X^2.$$
Hence,
\begin{equation}\begin{split}
\lefteqn{\cC(x)+\int_0^T\widetilde\lambda x_t(S^0_t+\gamma x_t)\,dt}\\
&=\frac\gamma2X^2+\int_0^TS^0_t\dot x_t\,dt+\eta\bigg(\int_0^T\dot x_t^2\,dt +\int_0^T( \lambda S^0_tx_t+\nu^2 x_t^2)\,dt \bigg).
\end{split}\label{GSriskMartEq2}
\end{equation}
The first two integrals on the right-hand side of \eqref{GSriskMartEq2} belong to $L^1(\bP)$ by \eqref{S0SquareIntEq} and \eqref{cXConditions}. Next, the fuel constraint $\int_0^T\dot x_t\,dt=-X$ and Jensen's inequality imply that
\begin{equation}\label{xt2EstimateEq}
x_t^2=\bigg(\int_t^T\dot x_s\,ds\bigg)^2\le (T-t)\int_0^T\dot x_s^2\,ds\in L^1(\bP).
\end{equation}
It follows that also the rightmost integral in  \eqref{GSriskMartEq2}, and in turn the entire expression \eqref{GSriskMartEq2}, belong to $L^1(\bP)$.

 Now suppose that $S^0$ is a martingale. When $\int x_t\,dS_t^0$ is a true martingale, our result follows by applying  integration by parts.  The martingale property of $\int x_t\,dS_t^0$, however, is not clear without additional integrability conditions.  We therefore proceed as follows.
 Since we have assumed that the filtered probability space $(\Om,\cF,(\cF_t),\bP)$ satisfies the usual conditions, the rightcontinuous martingale $(S^0_t)$ is  optional; see Theorem 65 in Chapter IV of \cite{DellacherieMeyerA}. Thus, the process $(S^0_{t\wedge T})$ is equal to the optional projection of the constant process $t\mapsto S_T^0$. Hence, 
$$\bE\Big[\,\int_0^TS^0_t\dot x_t\,dt\,\Big]=\bE\Big[\,\int_0^TS^0_t\,d x_t\,\Big]=\bE\Big[\,\int_0^TS^0_T\,dx_t\,\Big]=\bE[\,S^0_T(x_T-x_0)\,]=-S_0X,
$$
where in the second step we have used Theorem 57 in Chapter VI of \cite{DellacherieMeyerB} and the fact that  $\int_0^T|S^0_t\dot x_t|\,dt$ belongs to $L^1(\bP)$ due to our assumptions on $S^0$ and $x$.  This concludes the proof. \qed

\bigskip

\bigskip\noindent{\bf Proof of Theorem \ref{GenSemThm}:}  
Let $x\in\cX(T,X)$ be given.  Integrating by parts yields
\begin{equation}\label{YintbypartsEq}
\int_0^tS^0_s\dot x_s\,ds+\eta\int_0^t\lambda S^0_sx_s\,ds=S^0_tx_t-S^0_0x_0+\eta\int_0^tx_s\,dY_s.
\end{equation}
Hence we get from \eqref{GSriskMartEq2}  that 
$$\cC(x)+\int_0^T\widetilde\lambda x_t(S^0_t+\gamma x_t)\,dt=\frac\gamma2X^2-S_0X+\eta\bigg(\int_0^Tx_t\,dY_t+\int_0^T(\dot x_t^2+\nu^2x_t^2)\,dt  \bigg).
$$
Our problem therefore reduces to minimizing the expression
\begin{equation}\label{CostFunctionalOfStrategyEq}
\bE\Big[\,\int_0^Tx_t\,dY_t+\int_0^T(\dot x_t^2+\nu^2x_t^2)\,dt \,\Big]
\end{equation}
over $x\in\cX(T,X)$. We first consider the case $\nu>0$ and show that the optimal strategy is \eqref{Thm2OptStrategyEq} and that the minimal value of the cost functional \eqref{CostFunctionalOfStrategyEq} is
\begin{equation}\label{ReducedMinimalValueThm2Eq}
\begin{split}
\nu  X^2\coth(\nu  T)+\frac X{\sinh(\nu  T)}\bE\Big[\,\int_0^{T} \sinh(\nu (T-u) )\,dY_u\,\Big]\\-\frac14\bE\bigg[\,\int_0^T\Big(\frac{\bE\big[\,\int_{t}^{T}\sinh(\nu (T-u) )\,dY_u\,\big|\,\cF_t\,\big]}{\sinh(\nu  (T-t))}\Big)^2\,dt\bigg].
\end{split}\end{equation}

Let us study the various expressions in \eqref{ReducedMinimalValueThm2Eq}. Integrating by parts and using $Y_0=0$, we find that
$$\int_0^t\sinh(\nu(T-u))\,dY_u=\sinh(\nu(T-t))Y_t+\nu\int_0^t\cosh(\nu(T-u))Y_u\,du
$$
and hence that
\begin{equation}\label{SinhIntByParts}\begin{split}
\int_t^T\sinh(\nu(T-u))\,dY_u=-\sinh(\nu(T-t))Y_t+\nu\int_t^T\cosh(\nu(T-u))Y_u\,du.
\end{split}\end{equation}
The right-hand side is square-integrable due to \eqref{S0SquareIntEq}.

 Let
$M$ be a rightcontinuous version of the martingale 
$$\bE\Big[\,\int_0^{T}\sinh(\nu (T-u) )\,dY_u\,\big|\,\cF_t\,\Big]=\nu\bE\Big[\,\int_0^T\cosh(\nu(T-u))Y_u\,du\,\Big|\,\cF_t\,\Big],\qquad 0\le t\le T,
$$
which exists since the underlying probability space was assumed to satisfy the usual conditions. Clearly, $M$ is a square-integrable martingale.

Then
\begin{eqnarray*}
H_t&:=&\frac1{\sinh(\nu  (T-t))}\bE\Big[\,\int_{t}^{T}{\sinh(\nu (T-u) )}\,dY_u\,\big|\,\cF_t\,\Big]\\
&=& \frac1{\sinh(\nu  (T-t))}\Big(M_t-\int_0^{t}\sinh(\nu (T-u) )\,dY_u\Big)
\end{eqnarray*}
and so
\begin{equation}\label{dHtEq}
dH_t=\nu \coth(\nu  (T-t))H_t\,dt+\frac1{\sinh(\nu  (T-t))}\,dM_t -dY_t.
\end{equation}
Moreover, by \eqref{SinhIntByParts},
$$H_t=-Y_t+\frac\nu{\sinh(\nu  (T-t))}\int_t^T\cosh(\nu(T-u))\bE[\,Y_u\,|\,\cF_t\,]\,du,
$$
and so
\begin{equation}\label{}
|H_t|\le |Y_t|+\bE\big[\,\sup_{t\le u\le T}|Y_u|\,\big|\,\cF_t\,\big].
\end{equation}
We thus conclude from \eqref{S0SquareIntEq} that
\begin{equation}\label{supHtinL2Eq}
\sup_{0\le t\le T}|H_t| \in L^2(\bP).
\end{equation}
Therefore, we may define $N$ as  a rightcontinuous version of the martingale
$$-\frac14\bE\bigg[\,\int_0^TH_u^2\,du\,\Big|\,\cF_t\,\bigg],\qquad 0\le t\le T.
$$
We also define
\begin{eqnarray*}
G_t:=-\frac14\bE\bigg[\,\int_t^TH_u^2\,du\,\Big|\,\cF_t\,\bigg]=N_t+\frac14\int_0^tH_u^2\,du.
\end{eqnarray*}
It follows that
\begin{equation}\label{dGtEq}
dG_t=dN_t+\frac14 H_t^2\,dt.
\end{equation}

Now let  $x\in\cX(T,X)$ be given and define
$$C_t:=\nu x_t^2\coth(\nu (T-t))+{x_t}H_t+G_t.
$$
 When the theorem is correct,  $C_t$ should be  the minimal cost \eqref{CostFunctionalOfStrategyEq} for liquidating the position ${x_t}$ over $[t,T]$. Hence
$$\wt C_t:=\int_0^tx_s\,dY_s+\int_0^t(\dot x^2_s+\nu^2 x_s^2)\,ds+C_t 
$$
should be a submartingale for any $x\in\cX(T,{{X}})$ and a martingale for the optimal $x^*$. 

Let us first prove the following claim:
\begin{equation}\label{wtCtinL1Claim}
\sup_{0\le t\le T}|\wt C_t|\in L^1(\bP)\qquad\text{for all $x\in\cX(T,X)$.}
\end{equation}
 First, it follows from \eqref{YintbypartsEq}, \eqref{S0SquareIntEq}, and \eqref{xt2EstimateEq} that 
\begin{equation}\label{}
\sup_{0\le t\le T}\Big|\int_0^tx_s\,dY_s\Big|\in L^1(\bP).
\end{equation}
From \eqref{cXConditions} and \eqref{xt2EstimateEq} we get next that 
$$\int_0^T(\dot x^2_s+\nu  x_s^2)\,ds\in L^1(\bP).
$$
Furthermore, the fact that the function $t\mapsto t\coth t$ is bounded on $[0,\nu T]$ by a constant $c$ implies together with \eqref{xt2EstimateEq} that
$$\sup_{0\le t\le T}\nu x_t^2\coth(\nu (T-t))\le c\int_0^T\dot x_t^2\,dt\in L^1(\bP).
$$
Using again \eqref{xt2EstimateEq} and now \eqref{supHtinL2Eq} gives us 
$$\sup_{0\le t\le T}|x_tH_t|\in L^1(\bP).
$$
Finally, 
$$\sup_{0\le t\le T}|G_t|\in L^1(\bP)
$$
follows from \eqref{supHtinL2Eq} and Doob's $L^2$-maximal inequality. Our claim \eqref{wtCtinL1Claim} now follows by putting everything together. 

Using \eqref{dHtEq} and \eqref{dGtEq} we compute
\begin{eqnarray}d\wt C_t&=
&x_t\,dY_t+\bigg[ \dot x_t^2+\nu^2  x_t^2+2\dot x_tx_t\nu \coth(\nu (T-t))+\frac{\nu^2 x_t^2}{(\sinh(\nu(T-t)))^2}+\dot x_tH_t\nonumber\\
&&\qquad +x_t\nu \coth(\nu  (T-t))H_t+\frac14 H_t^2 \bigg]\,dt-x_t\,dY_t+\frac{x_t}{\sinh(\nu  (T-t))}\,dM_t+dN_t\nonumber\\
&=&\bigg[\dot x_t+x_t\nu \coth(\nu(T-t))+\frac12H_t\bigg]^2\,dt+\frac{x_t}{\sinh(\nu  (T-t))}\,dM_t+dN_t.\label{dwtCtEq}
\end{eqnarray}
Hence, $\wt C$ is the sum of a nondecreasing process and a local martingale.

Let us introduce the stopping times
$$\tau_n:=\inf\Big\{0\le t\le T\,\Big|\,|x_t|\ge n\Big\}\wedge T,
$$
with the convention $\inf\emptyset=+\infty$. Then 
$$\int_0^{t\wedge\tau_n}\frac{x_s}{\sinh(\nu  (T-s))}\,dM_s
$$
is a true martingale for $0\le t<T$ and $n\in\bN$. Therefore
$\bE[\,\wt C_{t\wedge\tau_n}\,]\ge \wt C_0=C_0
$.
By \eqref{wtCtinL1Claim} we may pass to the limit $t\ua T$ and $n\ua\infty$ and obtain
\begin{equation}\label{EwtCtIeq}
\bE\Big[\,\int_0^Tx_t\,dY_t+\int_0^T(\dot x_t^2+\nu^2x_t^2)\,dt \,\Big]=\bE\big[\,\wt C_T\,\big]\ge C_0. 
\end{equation}
Note that $C_0$ is equal to the asserted minimal value \eqref{ReducedMinimalValueThm2Eq} of our optimization problem.

Since $\bE[\,\wt C_{t\wedge\tau_n}\,]$ is nondecreasing in $t$ and $n$, we have an equality in \eqref{EwtCtIeq} if and only if $\bE[\,\wt C_{t\wedge\tau_n}\,]= \wt C_0$ for all  $t\in[0,T)$ and $n\in\bN$, which by 
\eqref{dwtCtEq} holds if and only if  $x$ satisfies the ODE
\begin{equation}\label{x*ODEEq}
\dot x_t=-x_t\nu \coth(\nu(T-t))-\frac12H_t,\qquad \text{$\bP$-a.s. for almost every $t\in(0,T)$.}
\end{equation}
When a solution to this equation exists in $\cX(T,X)$, it will be unique in this class by standard arguments. Thus, there is at most one optimal strategy.

Note that the  strategy \eqref{Thm2OptStrategyEq} can be written as
$$x^*_t
={\sinh\big(\nu (T-t)\big)}\bigg[\frac{X}{\sinh\big(\nu  T\big)}-\frac12\int_0^t\frac{H_s}{\sinh\big(\nu (T-s)\big)}\,ds\bigg]
$$
and hence solves \eqref{x*ODEEq}. We show now that $x^*\in\cX(T,X)$. First, we clearly have $x^*_0=X$. We show next that $x^*_t\to0$ as $t\ua T$. We have
\begin{equation}\label{CschIntegral}
\int_0^t\frac1{\sinh\big(\nu (T-s)\big)}\,ds=\frac1\nu\log\bigg(\frac{\sinh\big(\nu (T-t)\big)}{\sinh(\nu T)}\cdot\frac{\cosh(\nu T)-1}{\cosh\big(\nu (T-t)\big)-1} \bigg).
\end{equation}
As $t\ua T$, the right-hand side behaves like 
\begin{equation}\label{CschIntegralAsympt}
\frac1\nu\log \frac{\cosh(\nu T)-1}{\sinh(\nu T)}+\frac1\nu\log\frac1{2\nu(T-t)}.
\end{equation}
This function belongs to $L^1[0,T]$, and   with \eqref{supHtinL2Eq} we get $x^*_t\to0$ as $t\ua T$. We finally have to show that $\int_0^T\dot x_t^2\,dt\in L^1(\bP)$. We know that $x^*$ satisfies the ODE \eqref{x*ODEEq} and that $\int_0^TH_t^2\,dt\in L^1(\bP)$. Next,
$$\big|x^*_t\coth(\nu(T-t))\big|\le c_1+c_2\int_0^t\frac1{\sinh\big(\nu (T-s)\big)}\,ds\cdot\sup_{0\le s\le T}|H_s|^2
$$
for certain constants $c_1,c_2>0$. From \eqref{CschIntegral}, \eqref{CschIntegralAsympt}, and the fact that
$$\int_0^1\Big(\log\frac1{1-t}\Big)^k\,dt=k!
$$
for $k\in\bN$, we thus conclude that 
$$\int_0^T(\dot x_t^*)^2\,dt\le c_4+c_3\cdot\sup_{0\le s\le T}|H_s|^2\in L^1(\bP).
$$
Hence $x^*\in\cX(T,X)$, and the proof for  the case $\gamma>0$ is complete.

\bigskip

The case $\gamma=0$ can be analyzed by passing to the  limit $\nu\da0$. To this end, we note first that for $0\le t\le u\le T$
$$0\le \frac{\sinh(\nu(T-u))}{\sinh(\nu(T-t))}\le 1\qquad\text{and}\qquad \lim_{\nu\da0}\frac{\sinh(\nu(T-u))}{\sinh(\nu(T-t))}=\frac{T-u}{T-t}.
$$
It hence follows from the dominated convergence theorem for stochastic integrals \cite[p. 267]{Protter}
and \cite[Chapter V, Theorem 2]{Protter}
that 
$$\sup_{0\le t\le T}\bigg|\frac{\int_{t}^{T}\sinh(\nu (T-u) )\,dY_u}{\sinh(\nu  (T-t))}-\frac{\int_t^T(T-u)\,dY_u}{T-t}\bigg|\longrightarrow0\qquad\text{in $L^2(\bP)$ as $\nu\da0$.}
$$
Therefore, letting
$$x^\nu_t:={\sinh\big(\nu (T-t)\big)}\bigg[\frac{X}{\sinh\big(\nu  T\big)}-\frac12\int_0^t\frac{\bE\big[\,\int_{s}^{T}\sinh(\nu (T-u) )\,dY_u\,\big|\,\cF_s\,\big]}{\big(\sinh\big(\nu (T-s)\big)\big)^2}\,ds\bigg],
$$
we have $\sup_{0\le t\le T}|x^\nu_t-x^0_t|\to0$ in $L^2(\bP)$.
Passing to the limit $\nu\da0$ in \eqref{x*ODEEq} leads to the ODE
\begin{equation}\label{x0ODEEq}
\dot x_t=\frac{-x_t}{T-t}-\frac 1{2(T-t)}\bE\Big[\,\int_t^T(T-u)\,dY_u\,\Big|\,\cF_t\,\Big],\qquad \text{$\bP$-a.s. for almost every $t\in(0,T)$,}
\end{equation}
which is solved by $x^0$. Hence, we also have $\dot x^\nu_t\to\dot x^0_t$ in $dt\otimes dP$-measure. It follows that 
\begin{equation}\label{ValueFctConvergenceEq}
\liminf_{\nu\da0}\bE\Big[\,\int_0^Tx^\nu_t\,dY_t+\int_0^T\big((\dot x^\nu_t)^2+\nu^2(x^\nu_t)^2\big)\,dt \,\Big]\ge \bE\Big[\,\int_0^Tx^0_t\,dY_t+\int_0^T(\dot x^0_t)^2\,dt \,\Big].
\end{equation}
 For each fixed $\nu>0$, the expectation on the left-hand side is given by \eqref{ReducedMinimalValueThm2Eq}, and the latter expressions converge to
\begin{equation}\label{0ReducedMinimalValueThm2Eq}
\begin{split}
\frac{ X^2}T+\frac X{T}\bE\Big[\,\int_0^{T} (T-u) \,dY_u\,\Big]-\frac14\bE\bigg[\,\int_0^T\bigg(\frac{\bE\big[\,\int_{t}^{T}(T-u) \,dY_u\,\big|\,\cF_t\,\big]}{T-t}\bigg)^2\,dt\bigg],
\end{split}\end{equation} 
which is hence larger than or equal to $\bE\big[\,\int_0^Tx^0_t\,dY_t+\int_0^T(\dot x^0_t)^2\,dt \,\big]$. Finally, the optimality of $x^0$ follows from \eqref{ValueFctConvergenceEq} and the facts that each $x^\nu$ is optimal and that 
$$\bE\Big[\,\int_0^Tx_t\,dY_t+\int_0^T\big(\dot x_t^2+\nu^2x_t^2\big)\,dt \,\Big]\longrightarrow \bE\Big[\,\int_0^Tx_t\,dY_t+\int_0^T\dot x_t^2\,dt \,\Big]
$$
as $\nu\da0$ for each $x\in\cX(T,X)$.\qed

\bigskip

\noindent{\bf Proof of Theorem \ref{MartingaleThm}:} When $S^0$ is a martingale and $\nu>0$, 
\begin{eqnarray*}\bE\Big[\,\int_{t}^{T}\sinh(\nu (T-u) )\,dY_u\,\Big|\,\cF_t\,\Big]&=&\lambda \bE\Big[\,\int_{t}^{T}\sinh(\nu (T-u) )S^0_u\,du\,\Big|\,\cF_t\,\Big]\\
&=&\lambda S^0_t\int_{t}^{T}\sinh(\nu (T-u) )\,du\\
&=&\lambda S^0_t\frac1\nu\sinh(\nu (T-t) )\tanh\Big(\frac{\nu(T-t)}{2}\Big).
\end{eqnarray*}
Thus, \eqref{ReducedMinimalValueThm2Eq} reduces to \eqref{MartingaleMinimalCostEq} in the martingale case. Moreover, \eqref{Thm2OptStrategyEq} reduces to \eqref{GBMoptimalStrategyEq3}, because 
$$\frac{\tanh\big(\frac{\nu(T-t)}{2}\big)}{\sinh(\nu (T-t) )}=\frac1{1+\cosh(\nu(T-t))}.
$$
This concludes the proof of Theorem \ref{MartingaleThm}.\qed

\bibliography{Marketimpact}{}
\bibliographystyle{agsm}

\end{document}